\documentclass[a4paper,UKenglish,cleveref, autoref, thm-restate, authorcolumns]{lipics-v2019}

\graphicspath{{./figures/}}
\usepackage{booktabs}
\usepackage[titlenumbered, ruled, linesnumbered, vlined]{algorithm2e}
\usepackage{multirow}
\usepackage{todonotes}
\usepackage{soul}
\usepackage{hyperref}
\nolinenumbers
\let\oldnl\nl
\newcommand{\nonl}{\renewcommand{\nl}{\let\nl\oldnl}}

\def\fixme#1{\typeout{FIXED in page \thepage : {#1}}
\bgroup \color{red}{[FIXME: {#1}]} \egroup}

\title{Virtual Gang based Scheduling of Real-Time Tasks on Multicore Platforms}

\author{Waqar Ali}{University of Kansas\\{Lawrence, USA}}{wali@ku.edu}{}{}
\author{Rodolfo Pellizzoni}{University of Waterloo\\{Ontario, Canada}}{rpelliz@uwaterloo.ca}{}{}
\author{Heechul Yun}{University of Kansas\\{Lawrence, USA}}{heechul.yun@ku.edu}{}{}

\authorrunning{W. Ali and R. Pellizzoni and H. Yun} 

\Copyright{Waqar Ali and Rodolfo Pellizzoni and Heechul Yun} 



\category{} 

\relatedversion{} 

\supplement{}

\begin{document}

\maketitle

\begin{abstract}
We propose a virtual-gang based parallel real-time task scheduling approach for multicore platforms. Our approach is based on the notion of a virtual-gang, which is a  group of parallel real-time tasks that are statically linked and scheduled together by a gang scheduler. We present a light-weight intra-gang synchronization framework, called RTG-Sync, and virtual gang formation algorithms that provide strong temporal isolation and high real-time schedulability in scheduling real-time tasks on multicore. We evaluate our approach both analytically, with generated tasksets against state-of-the-art approaches, and empirically with a case-study involving real-world workloads on a real embedded multicore platform. The results show that our approach provides simple but powerful compositional analysis framework, achieves better analytic schedulability, especially when the effect of interference is considered, and is a practical solution for COTS multicore platforms.
\end{abstract}

\section{Introduction}\label{sec:intro}

High-performance multi-core based embedded computing platforms are increasingly being used in safety-critical real-time applications, such as avionics, robotics and autonomous vehicles. This is fueled by the need to efficiently process computationally demanding real-time workloads (e.g., AI and vision). For such workloads, multicore platforms provide ample opportunity for speedup by allowing these workloads to utilize multiple cores. However, the use of multicore platforms in safety-critical real-time applications brings significant challenges due to the difficulties encountered in guaranteeing predictable timing on these platforms.

In a multicore platform, tasks running concurrently can experience high timing variations due to shared hardware resource contention. The effect of contention highly depends on the underlying hardware architecture, which is generally optimized for average performance, thus often shows extremely poor worst-case behaviors~\cite{michael2019dos}. Furthermore, which tasks are co-scheduled at a time instance depends on the OS scheduler's decision and may vary over time. Therefore, to estimate a task's worst-case execution time (WCET) through  empirical measurements, which is a common industry practice, one might have to explore all feasible co-schedules of the entire taskset under a chosen OS scheduling policy. Any slight change in the schedule or change in any of the tasks can make ripple effects in the observed task execution times. In other words, timing of a task is \emph{coupled} with the rest of the tasks, the OS scheduling policy, and the underlying hardware.
For this reason, in the use-cases where hard real-time guarantees are a must, such as avionics, it is recommended to disable all but one core of a multicore processor~\cite{faa2016certification}, which obviously defeats the purpose of using multicore.

Gang scheduling was originally proposed in high-performance computing to maximize performance of parallel tasks~\cite{feitelson1992gang}, and many real-time varieties have been studied in the real-time community~\cite{berten2011rtns,KatoIshikawa,dong2017rtss_analysis,goossens2016optimal,goossens2010rtns_gangftp,goossens2016optimal,kato2009rtss_gangedf,wasly2019rtas_bundled}. In gang scheduling, threads of a parallel task are scheduled only when there are enough cores to schedule them all at the same time. Therefore, gang scheduling reduces scheduling induced timing variations and synchronization overhead~\cite{wasly2019rtas_bundled}. However, most prior gang scheduling studies do not consider the co-runner dependent timing variations due to contention in the shared hardware resources, and instead simply assume that WCETs already account for such effects, which may introduce severe pessimism in their analysis~\cite{yun2015ecrts}.

In our previous work~\cite{wali2019rtgang}, we presented \emph{RT-Gang} framework, which implements a restrictive form of gang scheduling policy for parallel real-time tasks, to address this problem of shared resource contention, by scheduling only one real-time gang task at a time, even if there are enough cores left to accommodate other real-time tasks~\footnote{In this paper, by a real-time task, we mean a \emph{periodically} activated task, which is composed of one or more parallel threads.}. This design philosophy of RT-Gang, although solves the problem of shared-resource contention among different real-time tasks, constrains the overall real-time schedulability of the system. Given that parallelization of a task  often does not scale well, and more cores are being integrated in modern multicore processors, it is unlikely to be a general solution for many systems. To mitigate this problem, we had introduced the notion of \emph{virtual gang}, which is defined as a group of real-time tasks that are always scheduled together as if they are members of a single parallel task. However, our previous work did not cover how such a virtual gang can be created, selected, and scheduled in ways to improve real-time schedulability.

In this paper, we rigorously examine the idea of a virtual gang and demonstrate that without careful consideration, the notion of virtual gang, as proposed in~\cite{wali2019rtgang}, can actually substantially \emph{decrease} the real-time schedulability of the system. In fact, we show that, in the worst-case, scheduling of a virtual gang task is equivalent of serializing all member tasks of the gang, which effectively nullifies any schedulability benefits of co-scheduling. We propose a light-weight intra-gang synchronization framework, which we call \emph{RTG-Sync}, to address the problem by ensuring all member tasks of a virtual gang are \emph{synchronously released}. It also provides easy to use APIs to create and destroy virtual gangs and their memberships. Moreover, RTG-Sync improves isolation between real-time tasks and best-effort tasks, and among the member tasks of each virtual gang, by integrating a page-coloring based last-level cache (LLC) partitioning mechanism.

RTG-Sync provides the following guarantees to the members of each virtual gang task: (1) all member tasks are statically determined and do not change over time; (2) no other real-time tasks can be co-scheduled; (3) best-effort tasks can be co-scheduled on any idle  cores, but their usage of LLC is restricted to a static partition via page-coloring so that they do not pollute the working-set of real-time task in the LLC. Moreover, maximum memory bandwidth usages of best-effort tasks are strictly regulated to a certain threshold value set by the virtual gang. These properties greatly simplify the process of determining task WCETs, because once a virtual gang is created, the other tasks that do not belong to the virtual gang cannot interfere with the member tasks, regardless of the OS scheduling policy, and the effect of shared hardware resource contention is strictly bounded. In short, RTG-Sync enables compositional timing analysis on multicore platforms\footnote{Timing analysis of a real-time system is considered compositional if analysis of a component can be carried out independently of other components.}.

We present virtual gang formation algorithms to help create virtual gangs and their member tasks from a given real-time taskset with the goal of maximizing system-level real-time schedulability. Since we are interested in demonstrating our technique on commercial-off-the-shelf platforms, we do not assume that a detailed hardware model is available; instead, we rely on measurement-based techniques for WCET estimation. Lastly, we describe how a classical single-core schedulability analysis can be applied to analyze the scheduling of parallel real-time tasksets on a multicore platform.

For evaluation, we first present schedulability analysis results with randomly generated parallel real-time tasksets. We then present case-study results conducted on a real multicore platform, demonstrating that the proposed framework achieves higher utilization and time predictability.

In summary, we make the following contributions:
\begin{itemize}
  \item We establish, with the help of concrete examples, the requirements for supporting the virtual gang abstraction in an operating system.
  \item We present RTG-Sync, a light-weight synchronization framework for ensuring   synchronous release of the member tasks for each virtual gang.
  \item We present virtual gang formation algorithms, which create virtual gangs and their   member tasks from a given taskset to improve system-level real-time schedulability. .
  \item  We implement our system in a real embedded multicore platform and evaluate our approach both analytically with generated tasksets and empirically with a case-study involving real-world workloads.\footnote{RTG-Sync and the analysis tools will be available as open-source.}
\end{itemize}

The rest of the paper is organized as follows. We provide necessary background in Section~\ref{sec:background}. In Section~\ref{sec:motivation}, we establish the requirements for  supporting virtual gangs in a gang-scheduling framework with the help of motivating examples. In Sections~\ref{sec:design} and~\ref{sec:gfa}, we explain the design of RTG-Sync and the gang formation algorithms respectively. In Section~\ref{sec:analysis},  we describe the schedulability analysis results using synthetic tasksets.  In Section~\ref{sec:eval}, we present our case-study evaluation results. We discuss related work in Section~\ref{sec:related} and conclude in  Section~\ref{sec:conclusion}.
\section{Background}\label{sec:background}
In this section, we provide necessary background.

\subsection{Rigid Real-Time Gang Task Model}\label{sec:model}
In this paper, we consider the \emph{rigid} real-time gang task model~\cite{goossens2010rtns_gangftp} with implicit deadlines, and focus on scheduling $n$ periodic real-time tasks, denoted by $T = \{\tau_1, \tau_2, ..., \tau_n\}$,  on a multicore platform with $m$ identical cores.

In the rigid gang model, each real-time task $\tau_i$ is characterized by three parameters $<h_i, C_i, T_i>$ where $h_i$ represents the number of cores (threads) required by the gang task to run, $C_i$ is the task's worst-case execution time (WCET), and $T_i$ represents its period, which is also equal to its deadline. The task model is said to be rigid because the number of cores  a task needs is fixed and does not change over time. The rigid gang task model is well suited for multi-threaded parallel applications, often implemented by using parallel programming frameworks such as OpenMP~\cite{dagum1998openmp}. We will review other task models in Section~\ref{sec:related}.

\begin{figure}
    \centering
    \includegraphics[height=5cm, width=0.8\linewidth]{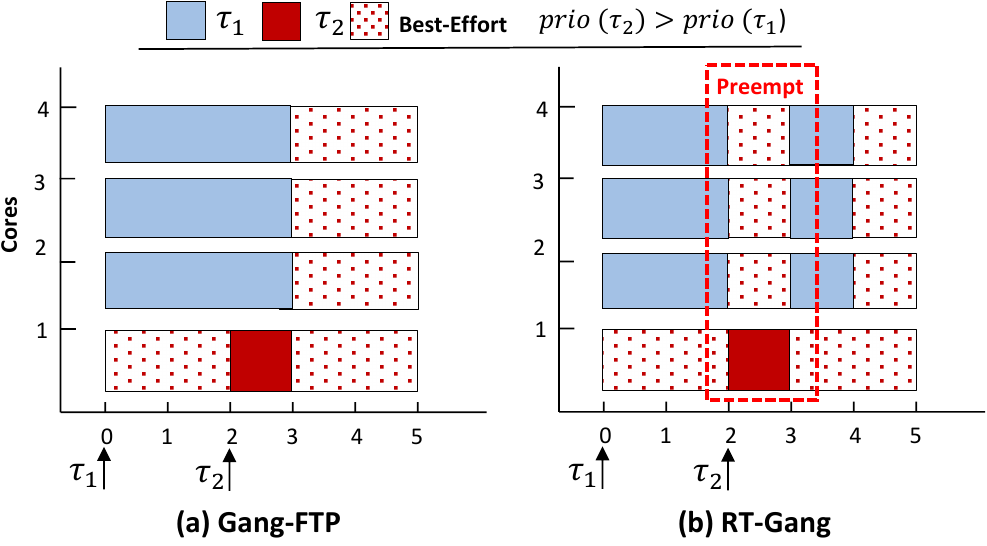}
    \caption{Illustration of the ``one-gang-at-a-time'' scheduling policy.}
    \label{fig:rtgang}
\end{figure}

\subsection{Priority-based Real-Time Gang Scheduling Algorithms}
Gang FTP~\cite{goossens2010rtns_gangftp} is a fixed-priority gang scheduling algorithm, which schedules  rigid and periodic real-time gang tasks as follows: At each scheduling event, the algorithm schedules the highest priority task $\tau_i$ on the first $h_i$ available cores (if exist) among the active (ready) tasks. The process repeats for the remainder of the active tasks on the remaining available cores. Gang EDF scheduler~\cite{lakshmanan2010scheduling} works similarly, except that the task's priority is not fixed but may dynamically change based on deadlines of the tasks at the moment (a task with more imminent deadline is given higher priority).

\subsection{RT-Gang}\label{subsec:rtgang}
RT-Gang is a recently proposed open-source real-time gang scheduler, which implements a restrictive form of Gang FTP scheduling policy in Linux kernel~\cite{wali2019rtgang}. According to this policy, at most one real-time task---which may be composed of one or more parallel threads---can be scheduled at any given time. When a real-time task is released, all of its threads are scheduled simultaneously on the available cores---if it is the highest priority real-time task---or none at all if there is a higher priority real-time task currently in execution.

Figure~\ref{fig:rtgang} shows the comparison of the ``one-gang-at-a-time'' scheduling policy of RT-Gang against Gang-FTP with a simple example, in which two real-time tasks $\tau_1$ and $\tau_2$ are scheduled on a multicore platform. Under Gang-FTP, $\tau_1$ and $\tau_2$ can be co-scheduled because their combined core requirement is equal to the total number of system cores. Under RT-Gang, such co-scheduling is not possible. All threads of $\tau_1$ are simultaneously preempted when the higher priority task $\tau_2$ arrives because real-time tasks must be executed one-at-a-time.

The rationale behind this ``simple'' gang scheduling policy---one-gang-at-a-time---is to eliminate the problem of shared hardware resource contention among co-executing real-time tasks by design. This also greatly simplifies schedulability analysis because it transforms the (complex) problem of multicore scheduling of real-time  tasks into the well-understood unicore scheduling problem. Since each real-time task is guaranteed temporal isolation, its worst-case execution time (WCET) can be tightly bounded as opposed to pessimistic estimation of WCET when co-scheduling of real-time tasks is allowed. Note that this restrictive gang scheduling is still strictly better than Federal Aviation Administration (FAA) recommended industry practice of disabling all but one cores~\cite{faa2016certification}.

\begin{figure}[t]
  \centering
  \includegraphics[height=5cm, width=0.8\linewidth]{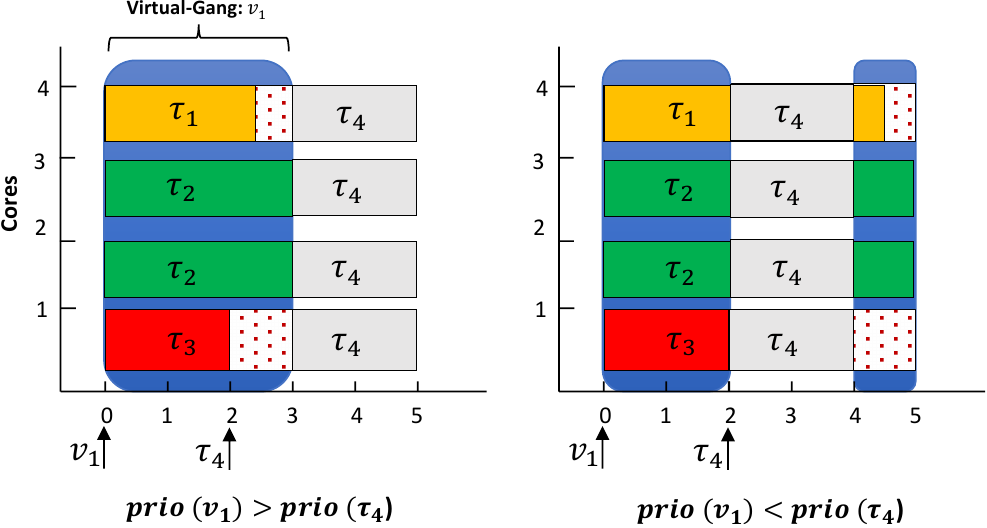}
  \caption{virtual gang concept. Adapted from~\cite{wali2019rtgang}}
  \label{fig:vgang}
\end{figure}

The obvious problem of low CPU utilization---because not all cores may be utilized by the scheduled one real-time gang task---is partially mitigated by  allowing co-scheduling of best-effort tasks on any idle cores, with a condition that their memory bandwidth usages are throttled to a certain threshold value, which is set by the real-time gang task, so that the impact of co-scheduling best-effort tasks to the critical real-time gang task is  strictly bounded. For throttling best-effort tasks, RT-Gang implements a hardware performance counter based kernel-level throttling mechanism~\cite{yun2013rtas}. Note, however, that although co-scheduling best-effort tasks can improve CPU utilization, it has no effects in improving low schedulability of real-time tasks under the strict one-gang-at-a-time policy.

\subsection{Virtual Gang}
To improve schedulability of real-time tasks under RT-Gang, we introduced the notion of virtual gang, which is defined as a group of real-time tasks that are explicitly linked and scheduled together as if they are the threads of a single real-time gang task under the scheduler. Figure~\ref{fig:vgang} illustrates the concept of a virtual gang, in which three separate real-time tasks, $\tau_1, \tau_2$ and $\tau_3$, form a virtual gang task $v_1$. The virtual gang task is then treated as single real-time gang task by the scheduler.

Unfortunately, however, the conditions under which virtual gangs can be created and how they can be effectively scheduled in ways to improve real-time schedulability of the system were not shown in our previous work~\cite{wali2019rtgang}.
\section{Motivation}\label{sec:motivation}
In this section, we show that the virtual gang concept, as described in~\cite{wali2019rtgang}, does not necessarily improve real-time schedulability of the system. We present two main challenges that need to be addressed for effective virtual gang based real-time scheduling.

\begin{figure}[t]
\centering
\begin{subfigure}[t]{0.32\textwidth}
    \centering
	\includegraphics[height=3.3cm, width=\textwidth]{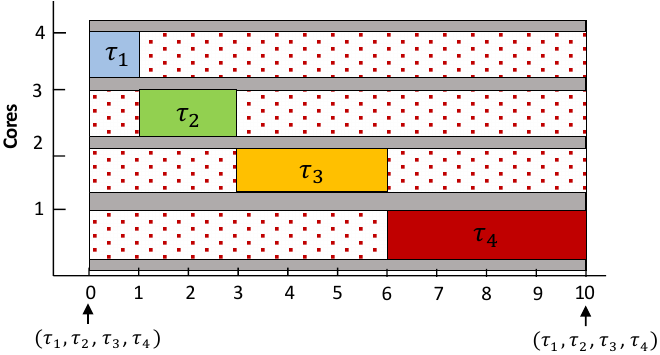}
	\caption{No virtual gangs}
	\label{fig:nos-rtgang}
\end{subfigure}
\hfill
\begin{subfigure}[t]{0.32\textwidth}
    \centering
	\includegraphics[height=3.5cm, width=\textwidth]{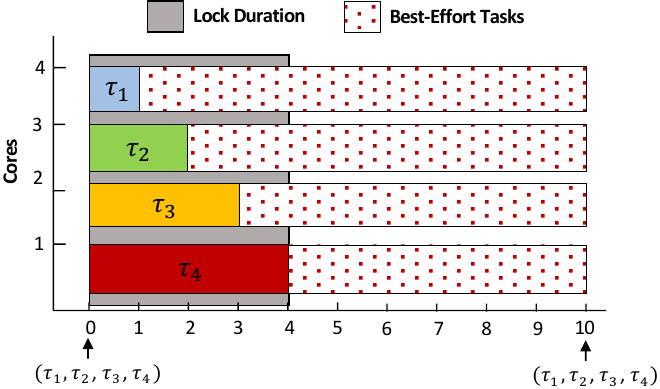}
	\caption{Synchronized virtual gang}
	\label{fig:nos-ideal}
\end{subfigure}
\hfill
\begin{subfigure}[t]{0.32\textwidth}
    \centering
	\includegraphics[height=3.3cm, width=\textwidth]{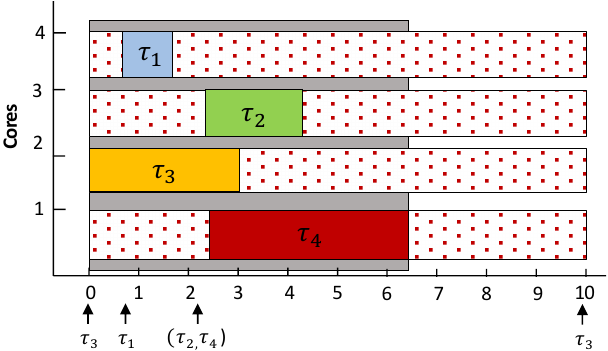}
	\caption{Unsynchronized virtual gang}
	\label{fig:nos-problem}
\end{subfigure}
\caption{Example schedules under different schemes}
\label{fig:nos}
\end{figure}

\subsection{Need of Synchronization}\label{sec:mot-nos}
The first major requirement for virtual gang is that all member tasks must have equal period and that they must be released synchronously.

As for the period requirement, if the linked member tasks of a virtual gang task do not  share the same period, then the consolidated gang task will not be effectively modeled as a single periodic task from the analysis point of view. Therefore, a virtual gang task can only be created when all of its member tasks have a common period.

\begin{table}[h]
	\centering
	\begin{tabularx}{\linewidth}{p{3cm}|p{3cm}|p{3cm}|p{3cm}}
		\toprule
		Task 			& WCET ($C$)	& Period ($T$)  & \# of Threads \\
		\midrule
		$\tau_{1}$		& 1 		& \multirow{4}{*}{10}	& \multirow{4}{*}{1}\\
		$\tau_{2}$		& 2 		& 			& \\
		$\tau_{3}$		& 3 		& 			& \\
		$\tau_{4}$		& 4 		& 			& \\
		\bottomrule
	\end{tabularx}
	\caption{Taskset parameters of the illustrative example}
	\label{tbl:tset1}
\end{table}

As for the synchronous release requirement, consider the taskset in the Table~\ref{tbl:tset1}, and  suppose it is scheduled on a quad-core platform (\texttt{m=4}). We consider scheduling of these tasks under the one-gang-at-a-time policy. When virtual gangs are not used, each task in the taskset executes as a gang by itself. This results in the scheduling timeline shown in the Figure~\ref{fig:nos-rtgang}. In this scheme, the completion time of the taskset is 10 time units, as the tasks execute sequentially one-at-a-time. Note that even though each of these tasks do not fully utilize the cores---use only one leaving three idle cores---the idle cores cannot be used (they are said to be ``locked'' by the gang scheduler) to schedule other real-time tasks and thus result in reduced real-time schedulability.

Now we consider the execution of this taskset as a virtual gang. Assuming that these are the only tasks that share the same period in our system and members of this taskset do not interfere with each other, an intuitive grouping of these tasks would be to run them at the same time across all four cores in the system. This results in the execution timeline shown in Figure~\ref{fig:nos-ideal}. In this scheme, the virtual gang completes in just 4 time units, after which other real-time tasks can be scheduled---i.e., improved real-time schedulability.

\begin{figure}[t]
    \centering
    \includegraphics[height=5cm, width=0.8\linewidth]{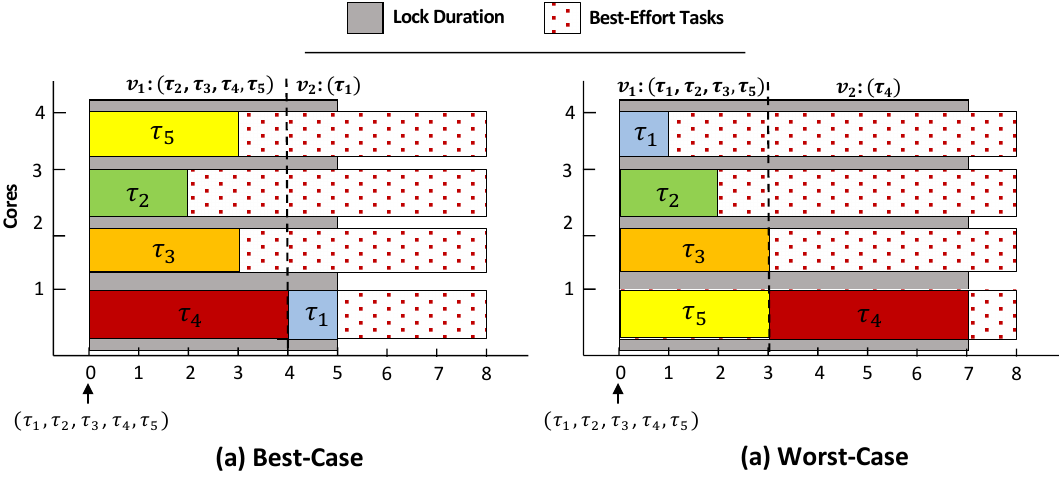}
    \caption{Example schedules under different gang formations.}
    \label{fig:gfp}
\end{figure}

However, the execution of the taskset in the virtual gang scheme assumes that the jobs of the members are perfectly aligned. If this is not the case, then the virtual gang task's execution time will be increased, as shown in Figure~\ref{fig:nos-problem}, and in the worst-case, it can be as bad as the original schedule without using virtual gang in terms of real-time task schedulability. 

\subsection{Gang Formation Problem}\label{sec:mot-gfa}
Another major challenge, when creating virtual gangs, is to decide which tasks to group together for concurrent execution.

Assume that in addition to tasks from Table~\ref{tbl:tset1}, there is one more task $\tau_5 = (h:1, C:3, P:10)$ which needs to be scheduled on the quad-core platform. In this case, it is required that the taskset be split into at least two virtual gangs since all five tasks in the taskset cannot execute simultaneously in our target system. Hence the problem is to find an optimal grouping of tasks into virtual gangs such that the execution time of the taskset is minimized. For the simple taskset considered here, it can be seen, with a little trial and error, that a virtual gang comprising $\tau_{2}$, $\tau_{3}$, $\tau_{4}$, $\tau_{5}$ and another one comprising just $\tau_{1}$ will achieve this goal, resulting in the execution timeline shown in the inset~(a) of Figure~\ref{fig:gfp}.

However if the tasks in a virtual gang are not carefully selected, the execution time of the taskset can increase significantly. In the example taskset, a virtual gang comprising $\tau_1$, $\tau_{2}$, $\tau_{3}$, $\tau_{5}$ and the other one comprising $\tau_{4}$ leads to an execution time of 7 time units as compared to 5 time units in the previous case; as can be seen in inset~(b) of Figure~\ref{fig:gfp}. 

Given a taskset, the problem of selecting the tasks which should be run together as virtual gangs so that the execution time of the entire taskset is minimized, is non-trivial. The problem is further complicated by the fact that the tasks in a virtual gang can interfere with each other when run concurrently due to shared hardware resource contention, which may require some degree of pessimism in estimating the virtual gang's WCET. 

Without taking the synchronization and gang formation problems into account, a strategy to improve system utilization via virtual gangs under RT-Gang may not lead to the desired results and may actually deteriorate the system's performance and real-time schedulability.
\section{RTG-Sync}\label{sec:design}
RTG-Sync is a software framework to enable virtual-gang based parallel real-time task scheduling on multicore platforms. 

As explained in the Section~\ref{sec:motivation}, synchronization between the members of each virtual gang is a key requirement for effective virtual gang based scheduling. For a typical multi-threaded process (task), synchronization between the threads of the process can be achieved by using a barrier mechanism available in the parallel programming library it uses (e.g., OpenMP barrier). However, such a barrier mechanism is tied to the particular  parallel programming framework, which is used by the particular parallel task, and is not designed to be used by disparate tasks for system-level scheduling. 

RTG-Sync provides a cross-process synchronization mechanism for virtual gangs, by utilizing existing OS-level inter-process communication (IPC) mechanisms. In addition, it provides an API to create and destroy virtual gangs and their membership. Furthermore, it integrates shared cache partitioning and memory bandwidth throttling mechanisms to bound the impact of interference in hardware resources. 

\begin{figure}[t]
	\centering
	\includegraphics[width=0.6\linewidth]{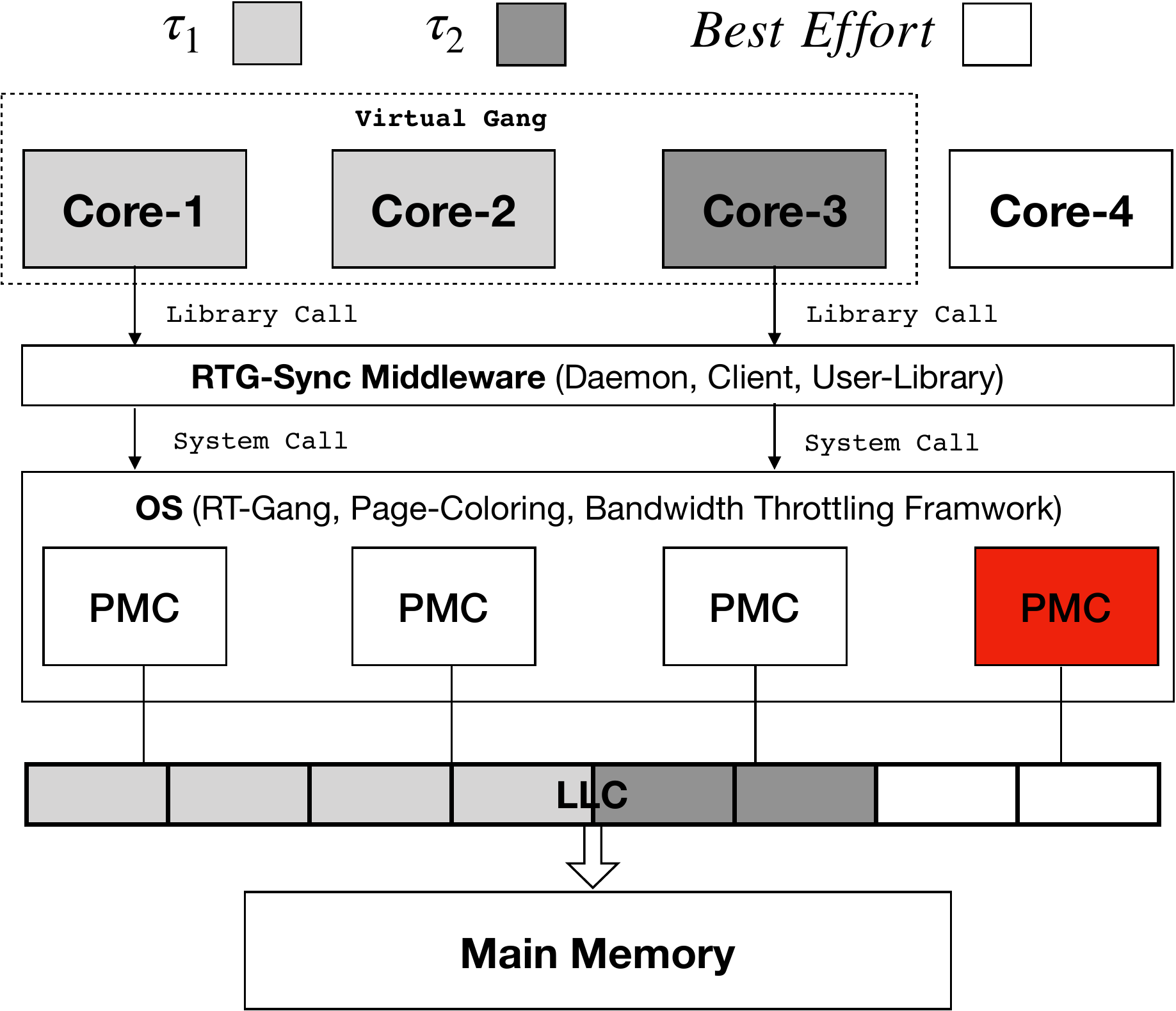}
	\caption{High level architecture of RTG-Sync framework. In this figure, $\tau_1$ and $\tau_2$ are real-time tasks and the resources allocated to them via RTG-Sync framework are color-coded.}
	\label{fig:arch}
\end{figure}

Figure~\ref{fig:arch} shows the high level architecture of RTG-Sync. The user-level component of RTG-Sync provides a specially designed system-wide barrier to each virtual gang so that all its member tasks can be synchronously released and scheduled by the kernel-level gang scheduler simultaneously. At the kernel-level, the modifications we have implemented ensure that the virtual gang execution is protected from interference by best-effort tasks through partitioning of LLC via page-coloring and memory level bandwidth throttling framework. In the figure, $\tau_1$ (on Core-1 and 2) and $\tau_2$ (on Core-3) are periodic real-time tasks of a virtual gang under RTG-Sync. The LLC has eight distinct partitions (colors), of which four are given to $\tau_1$, two are given to $\tau_2$, and the rest are reserved for best-effort tasks by RTG-Sync. In addition to the coloring based LLC partitioning, RTG-Sync additionally throttles the maximum memory bandwidth of Core-4, which can schedule best-effort tasks, to the virtual gang determined bandwidth thresholds so that the interference impact of best-effort tasks to the virtual-gang is bounded.

\subsection{Middleware}
RTG-Sync middleware consists of a server daemon and a client program. The primary service provided by the server is creating virtual gangs and initializing their associated resources. The server receives the number of processes, which need to be run as a single virtual gang, and creates a new memory mapped file in a predefined location, which is used for creating a system-wide barrier. When creating a virtual gang, we also specify the maximum memory bandwidth thresholds and LLC partitions (colors) for the best-effort cores (i.e., the cores that are not used to schedule the gang member tasks and thus can schedule any best-effort tasks). These parameters are enforced by the kernel-level mechanisms (Section~\ref{sec:kernel}) in order to bound the impact of co-scheduled best-effort tasks, if exist, to the virtual gang. A unique ID value is generated for each virtual gang, which is then used by the member tasks. Each member task shall make RTG-Sync user-library calls, to map the barrier file into its own address space and synchronize with other virtual gang members through the barrier.

The API call to register a process as a virtual gang member takes the virtual gang ID value issued by the RTG-Sync server along with the shared-resource requirement information for the calling process. Currently, each gang member task can specifies a portion of the LLC space, in terms of colors, the task is allowed to use. Internally, RTG-Sync makes a system-call to record the passed in parameters into the calling process’s task-structure in the kernel. Furthermore, it maps the system-wide barrier registered against the passed in virtual gang id into the calling process' address-space.

Once a process is registered as part of a virtual gang, the call to synchronize gang members is simple. It takes the barrier pointer returned by the  aforementioned API call and uses it to synchronize on the barrier. This call must be made by all member tasks at the start of their periodic execution. As soon as the waiter count for the barrier is reached, the member tasks are unblocked simultaneously; leading to desired alignment of their periodic execution. Because RTG-Sync requires all member tasks of a virtual gang to share the same period, no additional synchronization is necessary after they are simultaneously released at the beginning.

\subsection{Kernel Modification} \label{sec:kernel}
RTG-Sync uses the RT-Gang framework to provide gang scheduling inside the Linux kernel. We have made several changes to the RT-Gang framework. First, RT-Gang originally uses the SCHED\_FIFO priority value of a process to determine gang membership i.e., different processes which have the same SCHED\_FIFO priority are considered part of the same gang and are allowed simultaneous execution. There are two drawbacks of using this approach. First, it forces the RMS priority assignment in that tasks which have the same period must be executed as a single gang. Second, it makes the gang scheduler restricted to fixed priority assignment policy only.
In RTG-Sync, we have modified the kernel to instead use the virtual gang ID value recorded in the task's control block, to check gang membership, thus avoiding the aforementioned problems.

Second, we have integrated PALLOC~\cite{yun2014rtas}, which is a page-coloring framework, into RTG-Sync and modified it to allocate pages to a task based on the LLC color-map information stored in the task's control-block. We use PALLOC to perform two-level partitioning of the LLC. The first level statically partition's the LLC into two regions which are assigned to real-time tasks and best-effort tasks via Linux's Cgroups. The second level of partitioning is used to divide LLC between member tasks of a virtual gang as per the resource requirement information stored in the task's control-block.

Lastly, we have extended RT-Gang's memory bandwidth throttling framework to support separate read and write throttling capabilities~\cite{michael2019dos}. This can be used to improve performance of the best-effort tasks without impacting the virtual gang's performance. 

\section{Gang Formation Algorithm}\label{sec:gfa}
In this section, we describe the gang formation algorithm of RTG-Sync.

\subsection{Problem Statement}
For a given candidate-set of $N$ real-time rigid gang tasks with the same period $T$ and a given multicore platform with $m$ homogeneous CPU cores, we want to form a set of virtual gang tasks such that the total completion time of the virtual gangs is minimized. The member tasks of each virtual gang are synchronously released through the RTG-Sync framework. We further assume that for each task $\tau_k$ in the candidate-set, we can determine its WCET $C^*_k$ \emph{in isolation}, that is, without any co-running real-time or best-effort tasks. Consistent with employed industry practices, we do not restrict our methodology to any specific approach for computing the task WCET, i.e., either measurement or static analysis are acceptable. Note that while the task is executed without any co-runners, by definition $C^*_k$ must include the effects of synchronization and resource interference among concurrent threads of $\tau_k$.

We first present a brute-force algorithm to solve the virtual gang formation problem and then describe a heuristic based algorithm. Before delving into the  details of the algorithms, we first define key terms, which are used in the remainder of this section. 

\smallskip\noindent\textbf{System Configuration:} We define a system configuration ($G$), given a candidate-set, as a unique combination of virtual gangs which are sufficient to execute every task from the candidate-set. We use the following notation to denote a configuration: $ G_i = \{v_{i,1}, v_{i,2}, ..., v_{i,j}\} $ where each $v_{i,j} = (..., \tau_k, ...)$ denotes a virtual gang comprising tasks from the candidate-set. 

\smallskip\noindent\textbf{Completion Time:} The completion time of a configuration is defined as the time it takes for all the virtual gangs, which are part of the configuration, to complete their execution in a given period. Under our framework, the completion time of a configuration is equal to the sum of WCETs of the virtual gangs which are part of the configuration.

\subsection{Brute-Force Algorithm}
The brute-force algorithm for finding the best configuration from the candidate-set is stated in Algorithm~\ref{alg:gfa} and it revolves around the following key steps. Given the candidate-set, we generate all the possible configurations containing all possible pairings of tasks into virtual gangs (\texttt{line-4}). For this purpose, we write a recursive algorithm which, starting from the simplest configuration of tasks in the candidate-set into virtual gangs where every task runs as a gang by itself, successively generates more complex configurations by programmatically pairing tasks into viable\footnote{A virtual gang is viable if it requires up-to \texttt{m} cores to execute.} virtual gangs.

We first compute an estimated completion time of each configuration (\texttt{line-5}). For this step, we optimistically assume that the tasks inside a virtual gang do not interfere with each other or with best-effort tasks. Under this assumption, the execution time in isolation $C^*_{i,j}$ of each virtual gang $v_{i,j}$ is equal to the WCET of its longest running constituent task: $C^*_{i,j} = \max_{\tau_k \in v_{i,j}} \{ C^*_k\}$. Once the completion time of each configuration has been computed, all the configurations are ranked based on their completion time from shortest to longest (\texttt{lines-6:7}).  If two configurations have the same completion time, then the one which comprises smaller number of virtual gangs is given the higher rank. If a tie still exists between multiple configurations, then the one which is computed first by the algorithm is given the higher rank. The rank value is then used to sort the configurations from best to worst---best being one with the highest rank.

\begin{algorithm}[t]
\DontPrintSemicolon
\textbf{Input}: Candidate Set ($\mathbb{T}$), Number of Cores (\emph{m}) \\
\textbf{Output}: Best System Configuration \\
\nonl\;
\SetKwProg{funct}{function}{}{}
\SetKwFunction{func}{gang\_formation}
	\funct{\func{$\mathbb{T}$, m}}{
	configs = \emph{generate\_system\_configs} ($\mathbb{T}$, \emph{m})\\
	completionTimes = \emph{calc\_config\_times} (configs)\\
	rankedConfigs = \emph{rank\_configs} (configs, completionTimes)\\

	bestConfig = \emph{pick\_best\_config} (rankedConfigs)\\
	\nonl\;
	\While{True}{
		bestCompletionTime = bestConfig.completionTime\\
		\emph{add\_interference}(bestConfig)\\

		\uIf {(bestConfig.completionTime $\leq$ (1 + tolerance) * bestCompletionTime)}{
			\textbf{break}\\
		}
		\Else {
			completionTimes = \emph{update\_completion\_times} (bestConfig)\\
			rankedConfigs = \emph{rank\_configs} (configs, completionTimes)\\
			newBestConfig = \emph{pick\_best\_config} (rankedConfigs)\\

			\uIf {(newBestConfig == bestConfig)}{
				\textbf{break}\\
			}
			\Else {
				bestConfig = newBestConfig\\
			}
		}
	}
}
\KwRet{bestConfig}
\caption{Brute-Force Algorithm}
\label{alg:gfa}
\end{algorithm}

\textbf{Iterative Step}: Due to interfering effects over shared resources, the actual WCET $C_{i,j}$ of each virtual task $v_{i,j}$ might be (significantly) larger than its WCET in isolation $C^*_{i,j}$. For this reason, we need to update the completion time of the best configuration $G_i$ (\texttt{line-10}) by determining the WCET value $C_{i,j}$ of each virtual task in $G_i$. As mentioned before, we do not impose restrictions on how the WCET estimation is carried out. If a detailed model of the hardware platform and resource utilization of the tasks is available, then an analytical approach may be used to determine interference delay, as in~\cite{schoeberl2015t}. However, for commercial-off-the-shelf platforms, which are the target of this work, such approach is practically unfeasible. Therefore, we instead propose to empirically determine the WCET of each virtual task by executing it on the target platform under the synchronization framework of RTG-Sync, possibly in parallel with isolated best-effort tasks~\footnote{Based on the criticality level of the task, an additional safety margin might be added to the measurement-based WCET~\cite{jeoniso26262,hilderman2007avionics}.}. Independently of the WCET estimation method, based on the \emph{interference-updated} completion time of the configuration, there are two possibilities. If the completion time with interference is within a specified tolerance threshold (e.g., 20\%) of its computed value in isolation, the algorithm can finish and the best system configuration has been computed (\texttt{line-11}).

If, on the other hand, the completion time with interference is more than the tolerance threshold, the completion time of all configurations, which contain one or more of the virtual gangs from the best configuration and whose execution time changed in the interference evaluation, is recalculated and then the configurations are re-ranked (\texttt{lines-14:16}). If the best configuration stays the same, then the algorithm can finish (\texttt{line-17}). Otherwise, the iterative step is repeated with the new best configuration until the algorithm converges.

\smallskip\noindent\textbf{Complexity Analysis:} The worst-case complexity of the brute-force algorithm is linear with respect to the total number of system configurations due to the \texttt{while} loop on \texttt{line-8}. Given a candidate-set with $N$ tasks and a system with $m$ cores, the worst-case with respect to the number of unique system configurations arises when each task is single-threaded. In this case, the total number of system configurations is upper bounded by the following series sum:
\begin{align}\label{eq:sum}
&&|G| = \sum_{k=\lceil\frac{N}{m}\rceil}^{N}{S (N, k)}
\end{align}

where $S(N,k)$ is \emph{Stirling number of the second kind}~\cite{Graham1989}
and can be calculated using the following equation~\cite{Wolfram2019}:
\begin{align}\label{eq:fact}
&&S(N,k) = \frac{1}{k!}\sum_{i=0}^{k}{{(-1)^i}{k \choose i}{(k-i)^n}}
\end{align}

In the context of virtual gang formation, each $S(N,k)$ represents the unique number of ways to partition $N$ tasks into $k$ virtual gangs.

\begin{algorithm}[t]
\DontPrintSemicolon
\textbf{Input}: Candidate Set ($\mathbb{T}$), Number of Cores (\emph{m}) \\
\textbf{Output}: Taskset comprising virtual gangs \\
\nonl\;
\SetKwProg{funct}{function}{}{}
\SetKwFunction{func}{gang\_formation}

\funct{\func{$\mathbb{T}$, m}}{
	sortedTaskset = \emph{sort\_tasks\_by\_compute\_time} ($\mathbb{T}$) \\
	virtualGangSet = $[\ ]$ \\

	\nonl\;
	\While{not\_empty \emph{(sortedTaskset)}}{
		anchorTask = sortedTaskset.pop () \\
        \For{\emph{nextTask} in sortedTaskset}{
            \If{anchorTask.h + nextTask.h $\leq$ m}{
                sortedTaskset.remove (nextTask) \\
                anchorTask = \emph{create\_virtual\_gang} (anchorTask, nextTask) \\
            }
        }
        virtualGangSet.append (anchorTask) \\
	}
}
\KwRet{virtualGangSet}
\caption{Greedy Packing Heuristic}
\label{alg:gfah}
\end{algorithm}

\subsection{Greedy Packing Heuristic}
The brute-force algorithm is adequate when the candidate-set is small and the tasks in the candidate-set are heavily parallel. For lightly parallel tasks and large candidate-sets, the complexity of the brute-force algorithm rapidly becomes intractable. For this reason, we present a simple to use heuristic for gang formation as shown in Listing~\ref{alg:gfah}. The first step in the heuristic is to sort the tasks in the candidate-set in decreasing order of their WCETs in isolation $C_k^*$~(\texttt{line-4}). Then we remove the task with the highest WCET, which we call anchor task, and pack as many tasks with it for co-execution as permissible by the number of cores $m$ of the platform~(\texttt{lines-7:11}); giving preference to tasks with larger WCETs if multiple tasks can be paired with the anchor task. The tasks which are paired off are removed from the candidate-set. We continue this process until the candidate-set is empty~(\texttt{line-6}).

Once the virtual gangs are formed, we perform the interference evaluation, as described in the previous section, to empirically determine the WCETs $C_k$ of the virtual gangs under RTG-Sync synchronization framework. For each virtual gang, if the WCET $C_k$ is within an acceptable tolerance threshold (e.g., 20\%) of the WCET in isolation $C_k^*$, the virtual gang is accepted. If this is not the case then contrary to the brute-force algorithm, the virtual gang is rejected and its member tasks are considered separate gangs; there is no iterative step. The runtime of the heuristic is $\mathcal{O}(N^2)$ because of the loops on \texttt{line-6} and \texttt{line-8}.
\section{Schedulability Analysis}\label{sec:analysis}
As stated in Section~\ref{sec:background}, we consider the \emph{rigid} real-time gang task model~\cite{goossens2010rtns_gangftp} with implicit deadlines. We consider a set of $n$ periodic real-time tasks, denoted by $\tau = \{\tau_1, \tau_2, ..., \tau_n\}$, and a multicore platform with $m$ identical cores. Each task $\tau_i$ is characterized by three parameters $\tau_i = \{h_i, C_i, T_i\}$  where $h_i$ is the number of cores the task needs to run, $C_i$ is the WCET and $T_i$ is the period. For RTG-Sync, we assume that each $\tau_i$ represents a virtual gang, which may be created by a gang formation algorithm described in the previous section; note that in this case, $C_i$ is the updated WCET including inter-task interference effects.

Because the underlying gang scheduler schedules these virtual gang tasks one-at-a-time, the exact schedulability test for our system is a straight-forward  application of the standard unicore response time analysis under the rate-monotonic priority assignment scheme~\cite{Audsley93RTA}, as depicted in the following:
\begin{align}\label{eq:rta}
&&R_i^{n+1} = C_i + \sum_{\forall \tau_j \in hp(\tau_i)}{\Bigl\lceil\frac{R_i^n}{T_j}\Bigr\rceil C_j}
\end{align}.
The taskset is schedulable if the response time of every task is less than its period.

\subsection{Simulation Setup}\label{sec:simulation}
In this section, we present the schedulability results comparing RTG-Sync with other paralell real-time task scheduling approaches with synthetically generated parallel real-time tasksets. For analysis, we ignore best-effort tasks as we have means (bandwidth throttling and cache partitioning) to bound their impact to the real-time tasks (Section~\ref{sec:design}). We, however, do consider possible interference among the real-time tasks in the analysis as RTG-Sync does not protect bandwidth contention among the real-time tasks within a virtual gang.

For the real-time taskset generation, we first uniformly select a period $T_i$ in the range $[10, 1500]$. For each $T_i$, $N$ tasks $\tau_{i, j}$, where $N$ is randomly picked from the interval $[2, 5]$, are generated by selecting a WCET $C^*_{i, j}$ in the range $[T / 10, T / 5]$ and a parallelism level $h_{i, j}$. Note that $C^*_{i, j}$ represents the WCET of the task while running alone, as discussed in Section~\ref{sec:gfa}; the actual WCET $C_{i, j}$, including interference between co-scheduled tasks, depends on the employed scheduling policy, and we thus detail its computation later. The utilization $u_{i,j}$ of each $\tau_{i,j}$ is then calculated using the relation: $u_{i,j} = \frac{C^*_{i,j} \times h_{i,j}}{T_i}$.  If $u_{i,j}$ is less than the remaining utilization for the taskset, $C^*_{i,j}$ is adjusted so that $\tau_{i,j}$ fills the remaining utilization.  Otherwise, taskset generation continues until the desired level of utilization is reached.

\smallskip\noindent\textbf{Taskset Types:} Similarly to the work in~\cite{wasly2019rtas_bundled}, we consider three types of tasksets in our simulation, based on the allowed level of parallelization $h_{i,j}$ for the tasks in the taskset. For a \emph{lightly-parallel} taskset, $h_{i,j}$ is uniformly selected in the range $[1, \lceil{0.3 \times m}\rceil]$ ($m$ is the number of cores as defined earlier). For a \emph{heavily-parallel} taskset, the value of $h_{i,j}$ is picked from the range $[\lceil{0.3 \times m}\rceil, m]$. Finally, for \emph{mixed} taskset, the parallelization level $h_{i,j}$ is selected randomly from the interval $[1, m]$.

\smallskip\noindent\textbf{Priority Assignment:} We consider the rate-monotonic based priority assignment scheme: $prio (\tau_i) > prio (\tau_j)$ if $T_i < T_j$.  For tasks with the same period, we assign priorities based on task's WCET: $prio (\tau_{i,j}) > prio (\tau_{i,k})$ if $C_{i,j} < C_{i,k}$. 

\smallskip\noindent\textbf{Scheduling Policies:} For each taskset type, we calculate the schedulability results under four different scheduling policies. Under the \emph{RT-Gang} policy, the unicore response time analysis using Equation~\ref{eq:rta} is applied to calculate schedulability of the taskset under the one-gang-at-a-time scheduling. For \emph{RTG-Sync}, we first form virtual gangs from the given taskset and then use Equation~\ref{eq:rta} to calculate schedulability of the new taskset comprising the virtual gangs. Under \emph{RTG-Sync (BFC)}, virtual gangs are formed using the brute-force method whereas in \emph{RTG-Sync (GPC)}, the greedy packing heuristic is used to form virtual gangs. For the \emph{Gang-FTP} policy, we use the analysis in~\cite{wasly2019rtas_bundled} to calculate schedulability of the taskset under gang fixed-priority scheduling~\footnote{Note that the analysis in~\cite{wasly2019rtas_bundled} applies to a bundled gang model, but since the bundled model generalizes the rigid gang model, it still applies to our case.}. Concerning other schedulability analyses for the rigid gang model, we do not employ~\cite{KatoIshikawa,dong2017rtss_analysis} because they assume Gang EDF rather than FTP; nor we compare against~\cite{goossens2016optimal} as it requires creating static execution patterns over a hyper-period, which both significantly complicates the runtime and requires strictly periodic tasks. Finally, the \emph{Threaded} scheme models the scheduling of parallel tasks under vanilla Linux real-time scheduler, where the $m_{i,j}$ threads of each task $\tau_{i,j}$ are independently scheduled. In this case, we assess schedulability based on the state-of-the-art analysis for fixed-priority thread scheduling of DAG tasks in~\cite{fonseca2017rtns}; here $\tau_{i,j}$ is simply modeled as a DAG of $m_{i,j}$ nodes with the same execution time~\footnote{While we could also model the tasks according to the fork-join parallel model, as noted in~\cite{wasly2019rtas_bundled}, the state-of-the-art analysis for fork-join tasks would not perform better than~\cite{fonseca2017rtns}.}.

\smallskip\noindent\textbf{Interference Model:} We incorporate a simple interference model into our analysis to compute the WCET $C_{i,j}$ of each task mimicking shared resource interference between co-scheduled tasks on real platforms. For \emph{RT-Gang}, we simply set $C_{i,j} = C^*_{i,j}$, as under this policy, there is no co-scheduling among gangs by design. For all other policies, we consider both an \emph{ideal} case where $C_{i,j} = C^*_{i,j}$, and a more realistic case where $C_{i,j}$ is derived from $C^*_{i,j}$ based on the interference model.

According to this model, we randomly generate a resource-demand factor $r_{i, j}$ for each task $\tau_{i, j}$ in the range $[0, 1]$. Before applying the schedulability analysis, we calculate the worst-case resource utilization $R_{i,j}$ for each task, which is the sum of the demand of the task and the combined demand of the set of maximally resource intensive tasks which can get co-scheduled with the task under analysis as per the scheduling policy. We use $R_{i,j}$ to scale the WCET of the task as follows: $C_{i,j} = C^*_{i,j} \times \max(R_{i,j}, 1)$; intuitively, we assume that the task suffers no interference until the resource is over-utilized, after which we apply a linear scaling. In \emph{RTG-Sync (BFC)}, while calculating the completion time of each system configuration (\texttt{line-5} of Algorithm~\ref{alg:gfa}), the set of co-scheduled tasks is determined based on the virtual gangs and the completion time of each configuration is adjusted. In this case, the algorithm takes interference into account in selecting the optimal virtual gangs. In \emph{RTG-Sync (GPC)}, on the other hand, interference is taken into account, according to the strategy mentioned earlier, \emph{after} the best virtual gangs have been formed as per the heuristic.

For \emph{Gang-FTP}, we enumerate all possible sets of co-running tasks based on the remaining number of cores $M - m_{i,j}$, and pick the set with the maximal combined demand. For \emph{Threaded}, we assume that each independently scheduled thread of $\tau_{i,j}$ has a resource demand of $r_{i,j}/m_{i,j}$, and pick the $M - 1$ other threads (either of the same or different task) with maximal demands.

\subsection{Simulation Results}
{
\captionsetup{skip=0pt}
\begin{figure}[h]
    \includegraphics[height=6cm, width=\textwidth]{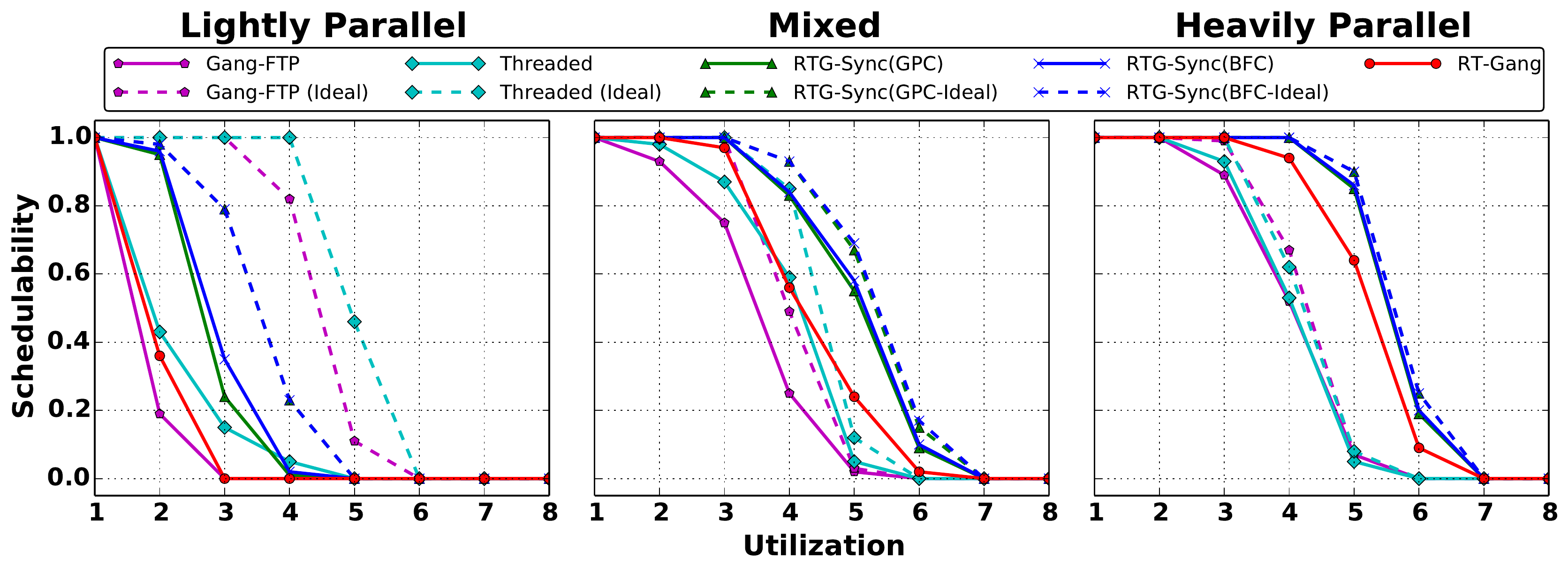}
    \caption{Schedulability results of the analyzed policies for different taskset types on 8 cores. Dashed lines are used when interference is not considered, while solid lines are used when it is incorporated in the analysis. For RT-Gang, because only one task can be scheduled at a time, interference cannot occur by design (thus no dashed lines).}
    \label{fig:sim-c8}
\end{figure}
}

Figure~\ref{fig:sim-c8} shows the schedulability plots for 8 cores ($m=8$) in three different types of tasksets---lightly parallel, mixed, and heavily parallel---depending on the degree of parallelism of the individual real-time tasks in the generated tasksets.

For lightly parallel tasksets, the \emph{Threaded} scheme gives the best schedulability results, followed closely by the \emph{Gang-FTP} policy, if interference model is not used (dashed lines). However, when interference is considered (solid lines), the schedulability under these policies deteriorates rapidly. As expected, \emph{RT-Gang} suffers the most for lightly parallel tasks as they under-utilize the cores. In comparison, both \emph{RTG-Sync (GPC)} and and \emph{RTG-Sync (BFC)} are significantly better than RT-Gang because the creation of virtual gangs improves utilization. Note that the effect of interference is less significant in RTG-Sync, compared to that of \emph{Gang-FTP} and \emph{Threaded}. This is due to the fact that under RTG-Sync, only the tasks of the same virtual gang can possibly interfere with each other, 
while a lot more tasks must be considered in Gang-FTP and Threaded.

For mixed and heavily parallel tasksets, both \emph{RTG-Sync} policies perform very similarly and outperform the rest regardless whether interference is considered or not. \emph{RT-Gang} improves considerably as a single parallel task can utilize more cores in the platform, though it still lags behind RTG-Sync. On the other hand, \emph{Gang-FTP} and \emph{Threaded} are significantly worse than RTG-Sync in both mixed and heavily parallel tasksets, and worse than RT-Gang in the heavily parallel tasksets. This can be attributed to the analysis pessimism needed to handle carry-in jobs in their schedulability analyses~\cite{wasly2019rtas_bundled,fonseca2017rtns}, which becomes more pronounced as the parallelism of the tasks increases. Because both RTG-Sync and RT-Gang use unicore fixed-priority schedulability techniques, they do not suffer from such analysis pessimism. 

Finally, in all cases, interference impact becomes less prominent as the parallelization of the taskset increases. This is because with highly parallel tasks, the opportunity of getting co-scheduled with other resource intensive tasks decreases, leading to improved schedulability.

\subsection{Effect of the Number of Tasks Per Period on RTG-Sync}
For RTG-Sync policies, the parameter $N$, which denotes the number of tasks generated for each period, is crucial. A high value of $N$ increases the possibilities of virtual gang formation and improves the schedulability under RTG-Sync policies. 

{
\captionsetup{skip=0pt}
\begin{figure}[h]
    \includegraphics[height=5cm, width=\textwidth]{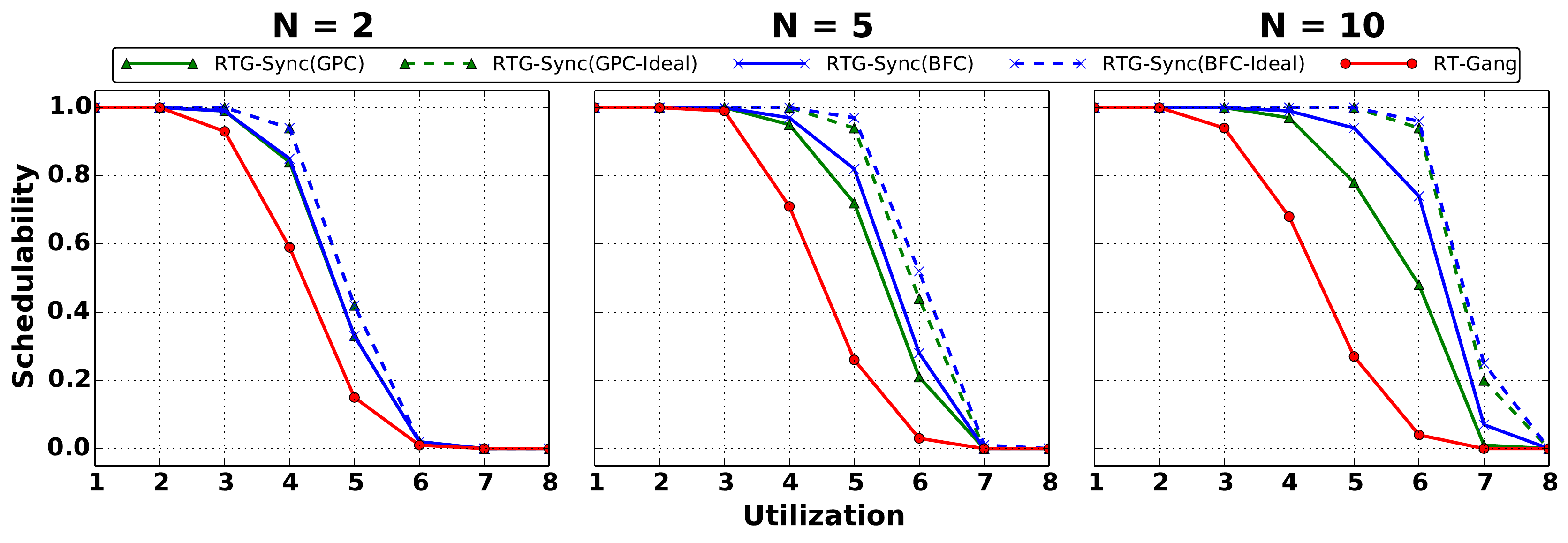}
    \caption{Effect of number of tasks per period ($N$) on RTG-Sync policies for mixed taskset}
    \label{fig:sim-c8-tpp}
\end{figure}
}

Figure~\ref{fig:sim-c8-tpp} shows the effect of $N$ on the performance of RTG-Sync policies~\footnote{For the sake of simplicity, we do not include the schedulability curves for \emph{Gang-FTP} and \emph{Threaded} because these policies are not fundamentally affected by changing the value of $N$.}. In this figure, the schedulability results for RTG-Sync policies are plotted for \emph{mixed} taskset types while statically changing the value of $N$ between the experiments while keeping the remaining setup the same as detailed in Sec~\ref{sec:simulation}. It can be seen that when $N = 2$, the schedulability curves under RTG-Sync policies are much closer to that for RT-Gang. The reason for this is that with $N = 2$, for each unique period $T$, only two tasks exist which can possibly form gang. However, the parallelism level of these tasks may be such that gang formation is not possible. With $N = 5$, the improvement in schedulability under RTG-Sync policies is more pronounced as compared to RT-Gang. Finally for $N = 10$, RTG-Sync policies provide significant improvement in schedulability over RT-Gang.

Another observation from this figure is that under the interference model, increasing the value of $N$ makes the difference in schedulability between the RTG-Sync brute-force algorithm and the heuristic more apparent. This indicates that as the possibilities of virtual gang formation increase, the heuristic becomes less likely to find the optimal virtual gang combination. This is to be expected since the heuristic does not take the interference among gang members into account during the gang formation process. Although it may be possible to design a more sophisticated heuristic for virtual gang formation, this would require incorporating an analytical interference model into the algorithm. As discussed in Section~\ref{sec:gfa}, this is highly challenging for commercial-off-the-shelf platform. Therefore, in this paper we preferred to demonstrate that we can still achieve significant improvements in schedulability with an algorithm that is interference-model agnostic; while we defer more complex, analytical methods for platforms where a detailed hardware model can be constructed to future work.
\section{Evaluation}\label{sec:eval}
In this section, we describe the evaluation results of RTG-Sync on a real multicore platform.

\subsection{Setup}\label{sec:ev-setup}
We use NVIDIA's Jetson TX-2~\cite{jetson-tx2} board for our evaluation experiments with RTG-Sync. The Jetson TX-2 board has a heterogeneous multicore cluster comprising six CPU cores (4 Cortex-A57 + 2 Denver\footnote{We do not use the Denver cores because of their lack of support for necessary hardware performance counters to implement the throttling mechanism.}). On the software side, we use the Linux kernel version 4.4 and patch it with the modified version of RT-Gang~\cite{rt-gang} to enable real-time gang scheduling at the kernel level along with best-effort task throttling and page-coloring frameworks, adding \texttt{$\sim$3000} lines to the architecture neutral part of the Linux kernel. In all our experiments, we put our evaluation platform in maximum performance mode which involves statically maximizing the CPU and memory bus clock frequencies and disabling the dynamic frequency scaling governor. We also turn off the GUI and networking components and lower the run-level of the system ($5 \rightarrow 3$) to keep the background system services to a minimum.

\subsection{Case-Study}\label{sec:ev-cstudy}
In this case-study, we demonstrate the effectiveness of using virtual gangs to improve system utilization, compared to the ``one-gang-at-a-time'' scheduling and Linux's default scheduler.

{
\renewcommand{\arraystretch}{1.25}
\begin{table}[h]
	\centering
	\begin{tabularx}{\linewidth}{p{2.35cm}|p{2.35cm}|p{2.35cm}|p{2.35cm}|p{2.3cm}}
		\toprule
		Task 			& WCET (ms)	&   Period
                (ms)	& \# of Threads 	& Priority \\
		\midrule
		$\tau_{BWT}^{RT}$	& 50.0 			    & 100.0	& 4		& 5   \\
		$\tau_{DNN-1}^{RT}$	& 8.2			    & 50.0	& 2		& 10  \\
		$\tau_{DNN-2}^{RT}$	& 8.2 			    & 50.0	& 2		& 10  \\
        	\hline
		$\tau_{cutcp}^{BE}$	& $\infty$		    & N/A	& 2		& N/A \\
		$\tau_{lbm}^{BE}$	& $\infty$		    & N/A	& 2		& N/A \\
		\bottomrule
	\end{tabularx}
	\caption{Taskset parameters for case-study}
	\label{tbl:case-study}
\end{table}
}

\begin{figure}[t]
	\centering
	\includegraphics[height=8cm, width=0.7\linewidth]{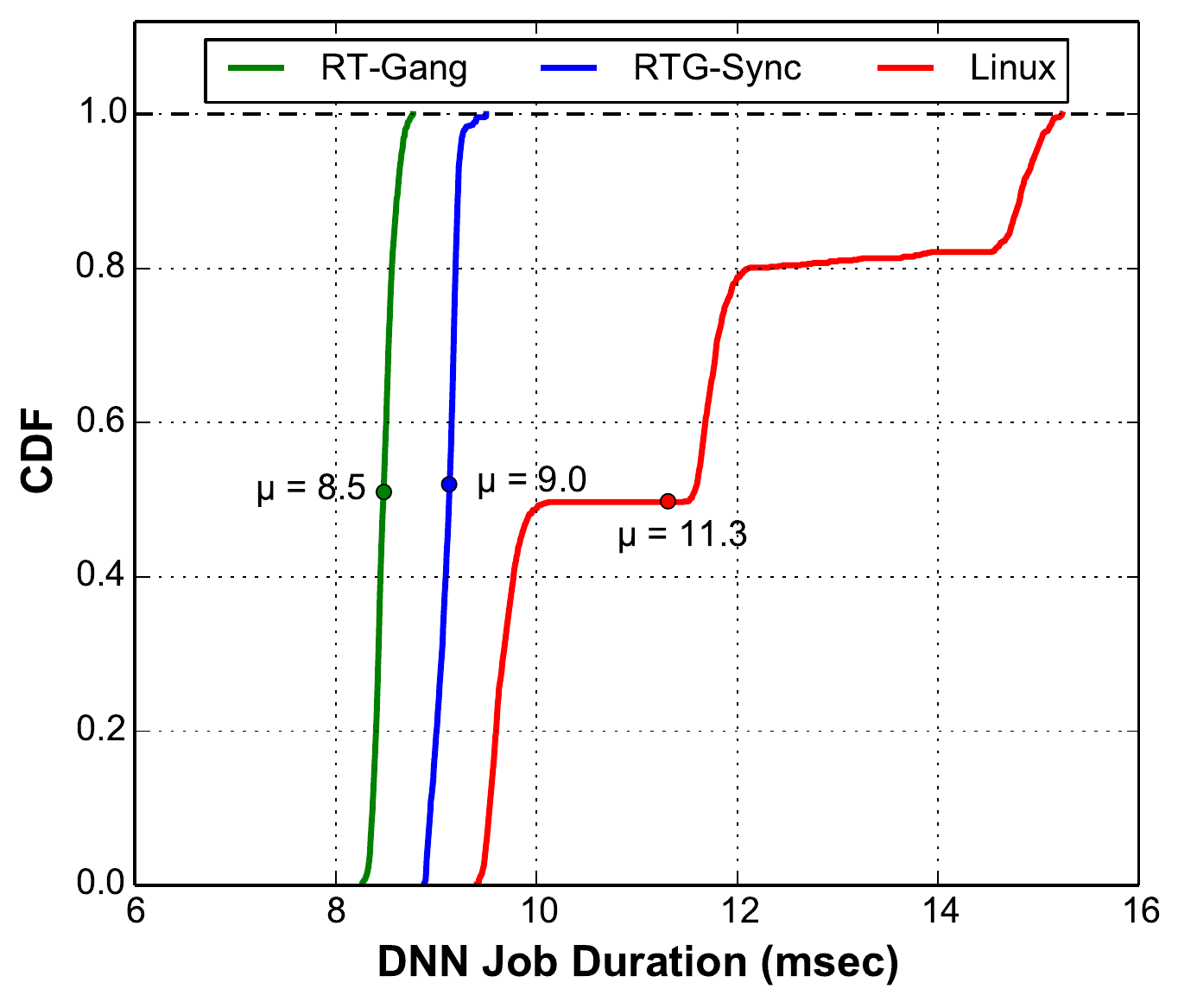}
	\caption{Distribution of job duration for $\tau_{DNN-1}^{RT}$}
	\label{fig:ev-cs-cdf}
\end{figure}

The taskset for the case-study is shown in Table~\ref{tbl:case-study}. It consists of three real-time tasks and two best-effort ones. For real-time tasks, we use the DNN workload from DeepPicar~\cite{michael2018deep} as two of the real-time tasks $\tau_{DNN-1}^{RT}$ and $\tau_{DNN-2}^{RT}$. Both DNN tasks use two threads each and have the same period of 50 ms. We use the synthetic bandwidth-rt benchmark as the third real-time task $\tau_{BWT}^{RT}$, which uses 4 threads and has a period of 100 ms. $\tau_{BWT}^{RT}$ is designed to be oblivious to shared resource interference but it creates significant shared hardware resource contention to DNN tasks under co-scheduling. As per the RMS priority assignment,  we assign higher real-time priority to DNN tasks than the bandwidth-rt task.

For best-effort tasks, we use two benchmarks from the Parboil benchmark suite~\cite{parboil}. Among the best-effort tasks, $\tau_{lbm}^{BE}$ is significantly more memory intensive than $\tau_{cutcp}^{BE}$. Both best-effort tasks use two threads each and are pinned to disjoint CPU cores.

We evaluate the performance of this taskset on Jetson TX-2 under three scenarios. The \emph{Linux} scenario represents the scheduling of the taskset under the vanilla Linux kernel. In \emph{RT-Gang} scheme, the real-time tasks are gang scheduled with the one-gang-at-a-time policy. Finally, under \emph{RTG-Sync}, we create a virtual gang, which is comprised of the two real-time DNN tasks. We assign 3/4th of the LLC to the virtual gang (two DNN tasks) and the rest 1/4th of the cache to the best-effort tasks. We do not, however, apply partitioning between the DNN tasks as sharing the cache space is beneficial in this case. 

Figure~\ref{fig:ev-cs-cdf} shows the cumulative distribution function of the job execution times of $\tau_{DNN-1}^{RT}$ under the three compared schemes. Note that this task has the highest real-time priority in our case-study. In this figure, the performance of $\tau_{DNN-1}^{RT}$ remains highly deterministic under both RT-Gang and RTG-Sync. In both cases, the observed WCET of $\tau_{DNN-1}^{RT}$ stays within \texttt{10\%} of its solo WCET---i.e., measured WCET in isolation---from Table~\ref{tbl:case-study}. However, under the baseline Linux kernel (denoted as \emph{Linux}), the job execution times of $\tau_{DNN-1}^{RT}$ vary significantly, with the observed WCET approaching \texttt{2X} of the solo WCET.

\begin{figure}[t]
\centering
\captionsetup[subfigure]{justification=centering}
    \begin{subfigure}[t]{\textwidth}
        \centering
	    \includegraphics[width=\linewidth]{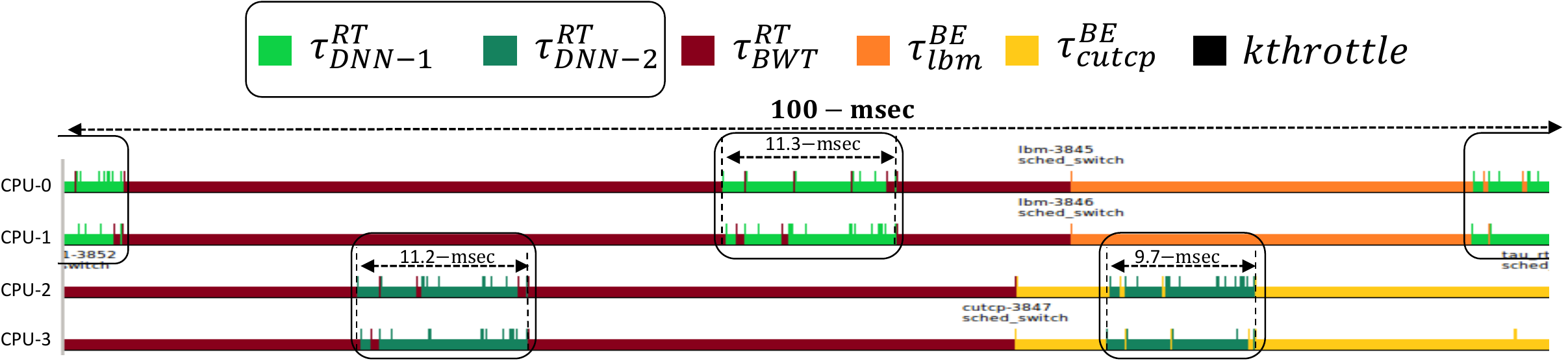}
	    \caption{Linux Default}
	    \label{fig:trc-global}
    \end{subfigure}
    \hfill
    \begin{subfigure}[t]{\textwidth}
        \centering
	    \includegraphics[width=\linewidth]{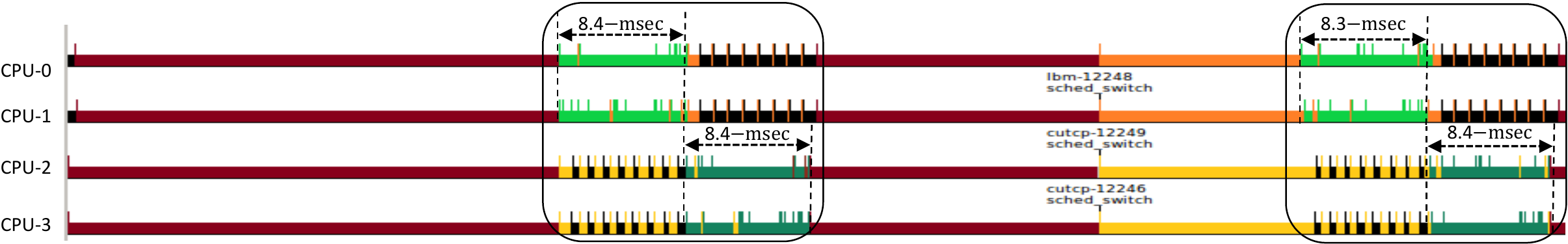}
	    \caption{RT-Gang}
	    \label{fig:trc-rtgang}
    \end{subfigure}
    \hfill
    \begin{subfigure}[t]{\textwidth}
        \centering
	    \includegraphics[width=\linewidth]{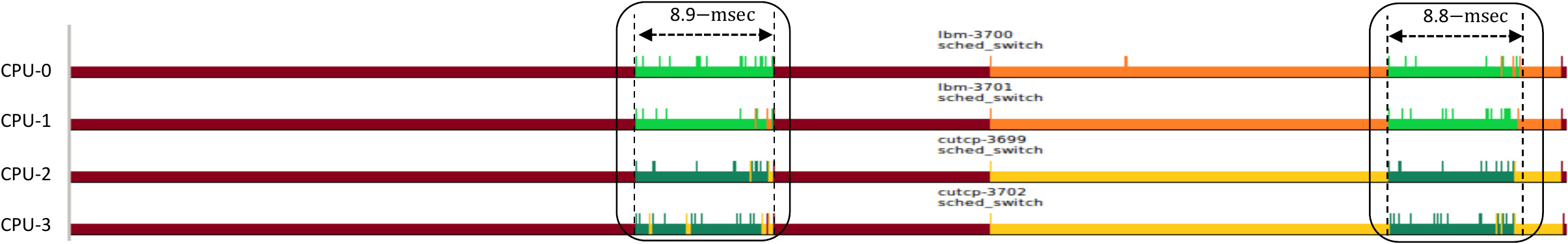}
	    \caption{RTG-Sync}
    	\label{fig:trc-rtgsynch}
    \end{subfigure}
\caption{Annotated KernelShark trace snapshots of case-study scenarios
  for one hyper period}
\label{fig:ev-cs-trc}
\end{figure}

The difference among the observed performance of $\tau_{DNN-1}^{RT}$ under the three scenarios can be better explained by analyzing the execution trace of the taskset in one hyper-period of 100 ms, which is shown in Figure~\ref{fig:ev-cs-trc}. Inset~\ref{fig:trc-global} displays the execution timeline under vanilla Linux. It can be seen that the DNN tasks suffer from two main sources of interference in this scenario. Whenever the execution of the DNN tasks overlaps with the execution of $\tau_{BWT}^{RT}$, the execution time of the task increases. The execution time also increases when the DNN tasks get co-scheduled with best-effort tasks. Note that the system is not regulated in any way in this scenario. Therefore, the effect of shared resource interference is difficult to predict, as evidenced in the CDF plot of Figure~\ref{fig:ev-cs-cdf}, which shows highly variable timing behavior. Under RT-Gang, on the other hand, the execution of DNN tasks is almost completely deterministic. Due to the restrictive one-gang-at-a-time scheduling policy, co-scheduling of DNN tasks with $\tau_{BWT}^{RT}$ is not possible. Moreover, the shared resource interference from the best-effort tasks is strictly regulated due to LLC partitioning and the kernel level throttling framework.

However, under RT-Gang, each DNN task executes as a separate gang by itself, which means that two system cores are left unusable for real-time tasks while the DNN tasks are executing because of the one-gang-at-a-time policy. This reduces the share  of total system utilization of the multicore platform, which can be used by other real-time tasks. Although the idle cores are utilized by best-effort tasks, the strict regulation imposed by DNN tasks means that the best-effort tasks are mostly throttled when they are co-scheduled with DNN tasks. Under RTG-Sync, both of these problems are solved by pairing $\tau_{DNN-1}^{RT}$ and $\tau_{DNN-2}^{RT}$ into a single virtual gang. In this case, the system is fully utilizable by real-time tasks. The execution of virtual DNN gang is completely deterministic due to the synchronization framework of RTG-Sync. Moreover, since there is no co-scheduling of best-effort tasks with real-time tasks, the throttling framework does not get activated and any slack duration left by real-time tasks can be utilized completely by best-effort tasks without imposing throttling.

\subsection{Overhead}\label{sec:overhead}
The runtime overhead due to RTG-Sync can be broken down into two parts. First is the overhead due to synchronization. This overhead is incurred only once during the setup phase of the real-time tasks which are members of the same virtual gang and it does not contribute to the WCETs of the periodic jobs. Second, the kernel level overhead is incurred due to the simultaneous scheduling of real-time tasks by the gang-scheduler. Since we use the RT-Gang framework for this purpose, the kernel level overhead is the same as reported in~\cite{wali2019rtgang}, which showed negligible overhead on a quad-core platform.
\section{Related Work}\label{sec:related}

Parallel real-time tasks are generally modeled using one of following three models: Fork-join model~\cite{lakshmanan2010scheduling,saifullah2013rtj_parallel, nelissen2012techniques,chwa2013global}, DAG model~\cite{baruah2012rtss_generalized,saifullah2014tpds_parallel,fonseca2017rtns} and gang task model~\cite{berten2011rtns,goossens2016optimal,goossens2010rtns_gangftp,kato2009rtss_gangedf,dong2017rtss_analysis,KatoIshikawa}. In the fork-join model, a task alternates between parallel (fork) and sequential (join) phases over time. In DAG model, which is a generalization of the fork-join model, a task is represented as a directed acyclic graph with a set of associated precedence constraints, which allows more flexible scheduling as long as the constraints are satisfied. Lastly, the gang model is further divided into three categories. Under the rigid gang task model~\cite{kato2009rtss_gangedf,dong2017rtss_analysis,goossens2016optimal}, the number of cores required by the gang are determined prior to scheduling and stay constant throughout its execution. In the moldable gang model~\cite{berten2011rtns}, the number of cores required by the gang are determined by the scheduler on a per-job basis but once determined, they are assumed to stay constant throughout the execution of the job. Finally, in the malleable gang model~\cite{goossens2008ipl_malleable}, the number of cores required by the job can change during the job's execution. Recently a bundled gang model is proposed in~\cite{wasly2019rtas_bundled}, which is a generalization of rigid gang model that allows more flexible parallel task modeling at the cost of increased analysis complexity. 

In the real-time systems community, fixed-priority and dynamic priority real-time versions of gang scheduling policies, namely Gang FTP and Gang EDF, respectively, are studied and analyzed~\cite{goossens2010rtns_gangftp,kato2009rtss_gangedf,dong2017rtss_analysis}. However, these prior real-time gang scheduling policies do not consider interference caused by shared hardware resources in multicore processors. On the other hand, the Isolation Scheduling model~\cite{huang2015isolation} and a recently proposed integrated modular avionic (IMA) scheduler design in~\cite{melani2017scheduling} consider shared resource interference and limit  co-scheduling to the tasks of the same criticality (in~\cite{huang2015isolation}) or those in the same IMA partition (in~\cite{melani2017scheduling}). However, they do not specifically target parallel real-time tasks and do not allow co-scheduling of best-effort tasks.  Also, to the best of our knowledge, all aforementioned real-time scheduling policies were not implemented in actual operating systems. Recently, a restrictive form of gang scheduling policy, which limits scheduling of just one gang task at a time, was proposed and implemented in Linux as a open-source project~\cite{wali2019rtgang,rt-gang}. The gang scheduler, called RT-Gang, provides strong temporal isolation by avoiding and bounding shared resource interference. However, it can significantly under-utilize computing resources in scheduling critical real-time tasks. Our work leverages the open-source RT-Gang scheduler and develops mechanisms and methodologies that improve real-time schedulability of the system at a marginal cost in terms of execution time predictability. 

Many researchers have attempted to make COTS multicore platforms more predictable with OS-level techniques.  A majority of prior works focused on \emph{partitioning} of shared resources among the tasks and cores to improve predictability.  Page coloring has long been studied to partition shared cache~\cite{liedtke97ospart,lin2008gaining,zhang2009towards,soares2008reducing,ding2011srm,ward2013ecrts,kim2013coordinated,ye2014coloris}, DRAM banks~\cite{yun2014rtas,liu2012software,suzuki2013coordinated}, and TLB~\cite{panchamukhi2015providing}. Mancuso et al.~\cite{mancuso2013rtas} and Kim et al.~\cite{kim2017attacking}, used both coloring and cache way partitioning~\cite{intelcat} for fine-grained cache partitioning. While these shared resource partitioning techniques can reduce space conflicts of some shared resources, hence beneficial for predictability, but they are often not enough to guarantee strong time predictability on COTS multicore platforms because of many undisclosed yet important shared hardware~\cite{valsan2016taming,michael2019dos}. Furthermore, partitioning techniques can lower performance and efficiency and are difficult to apply for parallel tasks.
\section{Conclusion}\label{sec:conclusion}

We presented a virtual gang based parallel real-time task scheduling approach for multicore platforms. Our approach is based on the notion of virtual gang, a group of parallel real-time tasks that are statically linked and scheduled together as a single scheduling entity. We presented an intra-gang synchronization framework and virtual gang formation algorithms that enable strong temporal isolation and high  real-time schedulability in scheduling parallel real-time tasks on COTS multicore platforms. We evaluated our approach both analytically and empirically on a real embedded multicore platform using real-world workloads. Our evaluation results showed the effectiveness and practicality of our approach. In future, we plan to extend our approach to support heterogeneous cores and accelerators such as GPUs.

\bibliography{main}

\end{document}